\documentclass[showpacs,aps]{revtex4}
\usepackage{amsmath,amssymb}
\begin{document}
\def\cs#1#2{#1_{\!{}_#2}}
\def\css#1#2#3{#1^{#2}_{\!{}_#3}}
\def\ket#1{|#1\rangle}
\def\bra#1{\langle#1|}
\def\expac#1{\langle#1\rangle}
\def\dbl{\hbox{${1\hskip -2.4pt{\rm l}}$}}
\def\bfh#1{\bf{\hat#1}}

\title{Disproof of Bell's Theorem: Reply to Critics}

\author{Joy Christian}

\email{joy.christian@wolfson.ox.ac.uk}

\affiliation{Perimeter Institute, 31 Caroline Street North, Waterloo, Ontario N2L 2Y5, Canada,}
\affiliation{Department of Physics, University of Oxford, Parks Road, Oxford OX1 3PU, England}

\begin{abstract}
This is a collection of my responses to the criticisms of my argument against the
impossibility proof of John Bell, which aims to undermine any conceivable local
realistic completion of quantum mechanics. I plan to periodically update this
preprint instead of creating a new one for each response.
\end{abstract}


\maketitle

\bigskip

{\underline{\bf Response \# 1 --- 26 March 2007}}:

\bigskip

In the preprint quant-ph/0703218, which appeared yesterday on this archives, my recent argument
to show that Bell's theorem fails for Clifford algebra valued variables has been criticized.
Apart from misrepresenting the central concepts employed in Ref.\cite{Christian} and thereby
misleading the reader, the main concern the critics raise is that, considering the opulence
of the geometrical quantities within Clifford algebra compared to vector algebra, it would be
impossible to recover the familiar scalar observables relevant to real experiments within my
Clifford-algebraic extension to the local realistic framework of Bell. More specifically, the
critics culminate their charge by declaring that, within my local realistic framework, it would
be impossible to derive ``a scalar in the RHS of the CHSH inequality. QED.''

\smallskip

If this were true, then it would certainly be a genuine worry. With hindsight, however, it would
have been perhaps better had I {\it not} left out as an exercise an explicit derivation of the
CHSH inequality in Ref.\cite{Christian}. Let me, therefore, try to rectify this pedagogical
deficiency here. Suppose we consider the familiar CHSH string of expectation values:
\begin{equation}
{\cal S}({\bf a},\,{\bf a'}\,,{\bf b}\,,{\bf b'})\,:=\,
{\cal E}({\bf a},\,{\bf b})\,+\,{\cal E}({\bf a},\,{\bf b'})\,+\,
{\cal E}({\bf a'},\,{\bf b})\,-\,{\cal E}({\bf a'},\,{\bf b'}).
\label{CHSH-op}
\end{equation}
As is well known, this string can be rewritten in terms of the products of the remote
observables, as
\begin{equation}
{\cal S}({\bf a},\,{\bf a'}\,,{\bf b}\,,{\bf b'})\,=\,
\int_{{\cal V}_3}\,[\;
A_{\bf a}(\boldsymbol\mu)\,B_{\bf b}(\boldsymbol\mu)\,+\,
A_{\bf a}(\boldsymbol\mu)\,B_{\bf b'}(\boldsymbol\mu)\,+\,
A_{\bf a'}(\boldsymbol\mu)\,B_{\bf b}(\boldsymbol\mu)\,-\,
A_{\bf a'}(\boldsymbol\mu)\,B_{\bf b'}(\boldsymbol\mu)\;]
\;\,d{\boldsymbol\rho}(\boldsymbol\mu). \label{probint}
\end{equation}
Now, using equations (16) and (17) of Ref.\cite{Christian} (and hence explicitly using my
Clifford algebra valued extension to the local realistic framework of Bell), the Clifford
product of any two of the above observables can be written as
\begin{equation}
A_{\bf a}(\boldsymbol\mu)\,B_{\bf b}(\boldsymbol\mu)\,=\,
(\,{\boldsymbol\mu}\cdot{\bf a})(\,{\boldsymbol\mu}\cdot{\bf b})\,=\,-\,{\bf a}{\bf b}\,
=\,-\,{\bf a}\cdot{\bf b}\,-\,{\boldsymbol\mu}\,({\bf a}\times{\bf b})\,\equiv\,
  -\,\exp\{{\boldsymbol\mu}\,{\bf z}\,\theta_{{\bf a}{\bf b}}\},\label{bi-product}
\end{equation}
where ${\theta_{{\bf a}{\bf b}}}$ is the angle from ${\bf a}$ to ${\bf b}$ about ${\bf z}$.
Here the last equivalence is elementary enough to derive, but can also be looked up in the
references cited in Ref.\cite{Christian}. Next, by substituting these values into equation
(\ref{probint}), we can rewrite it as 
\begin{equation}
{\cal S}({\bf a},\,{\bf a'}\,,{\bf b}\,,{\bf b'})\,=\,
-\int_{{\cal V}_3}\,[\;
\exp\{{\boldsymbol\mu}\,{\bf z}\,\theta_{{\bf a}{\bf b}}\}\,+\,
\exp\{{\boldsymbol\mu}\,{\bf z}\,\theta_{{\bf a}{\bf b'}}\}\,+\,
\exp\{{\boldsymbol\mu}\,{\bf z}\,\theta_{{\bf a'}{\bf b}}\}\,-\,
\exp\{{\boldsymbol\mu}\,{\bf z}\,\theta_{{\bf a'}{\bf b'}}\}\,]
\;\,d{\boldsymbol\rho}(\boldsymbol\mu). \label{reprobint}
\end{equation}
Finally, since the Euler identity ${\exp\{{\boldsymbol\mu}\,{\bf z}\,
\theta_{{\bf a}{\bf b}}\}\equiv\cos\theta_{{\bf a}{\bf b}}+{\boldsymbol\mu}\,{\bf z}\,
\sin\theta_{{\bf a}{\bf b}}}$ holds in Clifford
algebra, the directed measure ${d{\boldsymbol\rho}(\boldsymbol\mu)}$ is an isotropic probability
distribution, and the trivector ${{\boldsymbol\mu}=\pm\,I}$, averaging over the microstates
${\boldsymbol\mu}$ gives
\begin{equation}
{\cal S}({\bf a},\,{\bf a'}\,,{\bf b}\,,{\bf b'})\,=\,
-\,\cos\theta_{{\bf a}{\bf b}}\,-\,
\cos\theta_{{\bf a}{\bf b'}}\,-\,
\cos\theta_{{\bf a'}{\bf b}}\,+\,
\cos\theta_{{\bf a'}{\bf b'}}\,,
\label{My-CHSH}
\end{equation}
which is indeed ``a scalar in the RHS of the CHSH inequality'', and it is undoubtedly derived
from within the Clifford algebraic local realistic framework of Ref.\cite{Christian}. In fact,
it is well known that the numerical values of the RHS of the above equation for certain values
of the angles ${\theta_{{\bf a}{\bf b}}}$
lead to the celebrated violations of the CHSH inequality,
\begin{equation}
|{\cal S}({\bf a},\,{\bf a'}\,,{\bf b}\,,{\bf b'})|\,
\leqslant\,2.\label{eqCHSH}
\end{equation}
What is more, in {\it all} conceivable respects, our local realistic result (\ref{My-CHSH})
above is no different from what is usually predicted by using quantum mechanical operators,
and confirmed, of course, by real experiments.

\smallskip

Let me finally take this opportunity to thank Marek Czachor, Lucien Hardy, Lee Smolin, Ward
Struyve, and Chris Timpson for their valuable comments on, and insightful criticisms of,
Ref.\cite{Christian}.

\vfill\eject

{\underline{\bf Response \# 2 --- 11 July 2007}}:

\bigskip

Holman \cite{Holman} claims to have ``demonstrated'' that my local model for the EPR-Bohm
correlations ``runs into very serious difficulties'', while in the same breath acknowledging
that it actually reproduces the relevant predictions of quantum mechanics {\it exactly}.
In truth, however, Holman's ``demonstration'' is marred by several factual errors.

\smallskip

He begins with a complaint concerning the choice of my spin observables.
He suggests an alternative, ontologically sparser definition, which does not make
sense to me. For then why not simply keep
Bell's own vectorial observables? Since I have already discussed at length---in section
II of Ref.\cite{further}---the meaning of my observables, it would suffice for
me to simply say here that Holman is content with being an operationalist in this context,
whereas I am not.

\smallskip

The second issue of Holman's contention is the issue of sequential spin measurements. He
discusses the issue in the second column of the first page of his preprint, starting from
the previous column and spilling over considerably into the next, all the while resorting to
some hand-waving comparison between the quantum mechanical predictions and
those---supposedly---of my local model. What he fails to realize, however, is that his whole
discussion is marred by vector-algebraic prejudices. In particular, he disregards the fact that
the basis vectors in my Clifford-algebraic model are subject to non-commutative product
relations. Consequently, his vector-algebraic hand-waving has no relevance for my model. In
any case, the entire discussion can be summed up in a very simple question: Does the Malus's
law for sequential spin measurements hold in my model? The answer is `yes', and the derivation
can be found in Ref.\cite{further}.

\smallskip

To ``demonstrate'' next of his ``serious difficulties'' with my model, Holman begins by
asserting that any local realistic observables must satisfy the relation
\begin{equation}
A_{\bf n}(\lambda)\,=\,-\,
B_{\bf n}(\lambda)\,,
\label{bell-c1}
\end{equation}
on empirical grounds, and since (according to him) my observables do not seem to satisfy
this relation, my model contradicts experimental facts. This claim, of course, is simply
false. Again, what Holman does not seem to realize is that the above relation is an assumption
that Bell had to make in order to satisfy the perfect correlation constraint,
\begin{equation}
{\cal E}({\bf n},\,{\bf n}) = -\,1,
\label{bell-c2}
\end{equation}
which was adapted by EPR from quantum mechanics. The constraint any local realistic model
must satisfy is not (\ref{bell-c1}), but (\ref{bell-c2}), which is indeed satisfied by my model.
This is a very strong constraint, and---limited by the commutativity of his
variables---Bell had no choice but to assume relation (\ref{bell-c1}) to satisfy it, in order
to derive his inequality (see, e.g., section 3.1 of Ref.\cite{Clauser-Shimony}, or section II of
Ref.\cite{GHSZ}). The failure to appreciate this elementary fact has led Holman to make some
fallacious claims about my model, despite the fact that it reproduces all of the relevant
empirical predictions of quantum mechanics exactly \cite{further}.
His remaining discussion, therefore, is as devoid of merit as his conclusions are.

\smallskip

Finally, Holman concludes his almost entirely dubious analysis with a speculation: It
is highly unlikely that any future theory of physics would be locally causal.
Such an attitude actually reminds me of one of the most eminent philosophers of all
time: Immanuel Kant. After the spectacular successes of Newton's theory of gravity, Kant
endorsed Newtonian action-at-a-distance by arguing that it was no more problematic than
action by means of a physical contact, since in both cases a material body is simply acting
outside itself. Such a shift in attitude towards action-at-a-distance is well taunted by Mach:
The arrival of Newton's theory disturbed almost everyone because it seemed to be based on
an ``uncommon unintelligibility.'' In due course, however, and with sufficient empirical
support, it no longer disturbed anyone because it had become ``common unintelligibility''.
Similarly, it is perhaps not unfair to note today that the familiarity and successes of
quantum theory have reduced quantum nonlocality to a ``common unintelligibility'', not the
least because of the spectacular vindications of the theory in the experimental
tests of Bell inequalities. But who is to say that these tests are not, instead, vindicating
my own little local realistic model? After all, the predictions of my model for these tests
are no different in any respect whatsoever from those based on quantum mechanics!

\bigskip

{\underline{\bf Response \# 3 --- 17 July 2007}}:

\bigskip

The main issue Philippe Grangier has raised is indeed worthy of proper understanding, especially
to avoid the kind of confusions that he has exemplified in his critique \cite{Grangier}. To my
mind, however, I have already addressed this issue in Ref.\cite{further}. It is very much tied
up with the meaning of the word ``bivector'' within Clifford algebras. Rather surprisingly, this
word has proven to be an obstacle to the message I have been trying to put across. It would be a
great pity if a mental block against this word is allowed to obscure the truth. To prevent this
from happening, let me once again explain what is meant by a bivector, and then show that the
concerns of Philippe Grangier are grossly misplaced.

\smallskip

Before I begin, however, I must point out that the summary of my model in the beginning of
Grangier's preprint is at best a caricature. And the quotation of one-half of one of my
sentences in the second of his paragraphs is not only out of context, but has been exploited
by him in a misleading manner. In particular, I am certainly not holding my breath until ``a
yet to be discovered'' theory turns up so that I can extract the binary outcomes from
my observables, thus completing my model. In fact, the quotation of my words in his footnote [4]
is also completely discordant with what I have said. The reader, therefore, is urged to consult
Ref.\cite{further}, to find out what I have actually said, and within what context. My overall
impression of his critique is that he has formed a prior opinion of my model, without actually
trying to properly digest it first. Sadly, this seems to be the emerging pattern among all my
critics.

\smallskip

The essence of Grangier's worry is this: how are the binary outcomes ${\pm 1}$ extracted
from my observables, which are supposed to be bivectors. Actually, the very fact that he
has raised this worry reveals that---despite my efforts in Ref.\cite{further}---he has failed
to understand the meaning of my observables. To be sure, it is difficult to put aside our
vector-algebraic upbringing and embrace the unfamiliar Clifford algebraic notions, but that
is exactly what one must do to understand the nature of my observables. For example, in Clifford
algebra division by a vector is perfectly kosher, whereas in vector algebra it has no meaning.
More relevantly for us, a bivector in Clifford algebra must not be thought of as just ``two''
vectors, but as an entity of its own. And just as other elements of the algebra
${Cl_{3,0}\,}$, a bivector is {\it completely} characterized by three attributes, and three
attributes {\it only}. These are: (1) its magnitude,
(2) its direction, and (3) its sense of rotation. The magnitude of a bivector is determined
simply by it norm, as I have done in Ref.\cite{further}, and it is universally
fixed to be unity for the bivectors of my model. Moreover, its direction is determined by the
direction of the spin analyzer itself, so it too is fixed for a given analyzer. Thus, the only
attribute that
can possibly be revealed in any experiment is its sign, ${+}$ or ${-}$, which corresponds to the
sense of its rotation. In other words, the issue of ``a way to extract the ``sign'''' that
Grangier is worried about, is simply nonexistent in my model. To understand this better,
it is worth comparing our scenario with that in Bell's local model. In Bell's model the hidden
variable is a unit vector ${\boldsymbol\lambda}$, which, when projected onto the
direction of a spin analyzer ${\bf n}$, gives ${{\boldsymbol\lambda}\cdot{\bf n}}$. Now this
quantity is more than just a sign, and hence Bell uses the sign function to extract
the ${+}$ or ${-}$ out of it. Thus he takes his observables to be
${sign({\boldsymbol\lambda}\cdot{\bf n})}$. As we just saw,
no such {\it ad hoc} procedure is necessary in my model. 

\smallskip

Another way to appreciate the above comments is to recall that bivectors are actually nothing
but quaternionic numbers in disguise. The emphasis here is on the word {\it numbers}. Now,
a single quaternion by itself cannot be said to be either commuting or non-commuting, since the
concept of an algebra is a relational concept. One must have more than one quantity to form an
algebra. Thus, as far as an outcome of a single measurement is concerned, the fact that the
corresponding observable may be a part of a non-commuting algebra is of no relevance. It is only
when this observable is compared with another in some mathematical operation---such
as a product---the notion of commutativity or non-commutativity makes sense. This fact,
actually, brings us to the next of Grangier's worries.

\smallskip

Amusingly enough, after initially raising the high alert on how the signs of the observables are
extracted in my model, Grangier makes a complete u-turn by acknowledging the opposite: ``In other
words, though it seems that the only ``available'' information in the algebraic quantities
${{\boldsymbol\mu}\cdot{\bf a}}$ and ${{\boldsymbol\mu}\cdot{\bf b}}$ are their signs, ...''
Yes! That is all there is to it! But, alas, he then continues his sentence: ``...averaging over
non-commuting algebraic quantities is certainly not equivalent to averaging over commuting
real functions.'' Well, let us find out whether this is actually true. Suppose we average
over the product of the observables
${{\boldsymbol\mu}\cdot{\bf a}}$ and ${{\boldsymbol\mu}\cdot{\bf b}}$.
As in equation (19) of Ref.\cite{Christian}, what we obtain is:
\begin{equation}
\int_{{\cal V}_3}(\,{\boldsymbol\mu}\cdot{\bf a}\,)
(\,{\boldsymbol\mu}\cdot{\bf b}\,)\;\,d{\boldsymbol\rho}({\boldsymbol\mu})
=\,-\,{\bf a}\cdot{\bf b}\,-\int_{{\cal V}_3}
{\boldsymbol\mu}\cdot({\bf a}\times{\bf b})\;\,d{\boldsymbol\rho}({\boldsymbol\mu})
=\,-\,{\bf a}\cdot{\bf b}\,+\,0\,=\,-\,\cos\theta_{{\bf a}{\bf b}}\,.\label{derive-1}
\end{equation}
Suppose now we reverse the order of ${{\boldsymbol\mu}\cdot{\bf a}}$ and
${{\boldsymbol\mu}\cdot{\bf b}}$ to see whether the non-commutativity of these observables
leads to any difference in behavior compared to what one would expect from any commuting real
functions:
\begin{equation}
\int_{{\cal V}_3}(\,{\boldsymbol\mu}\cdot{\bf b}\,)
(\,{\boldsymbol\mu}\cdot{\bf a}\,)\;\,d{\boldsymbol\rho}({\boldsymbol\mu})
=\,-\,{\bf b}\cdot{\bf a}\,-\int_{{\cal V}_3}
{\boldsymbol\mu}\cdot({\bf b}\times{\bf a})\;\,d{\boldsymbol\rho}({\boldsymbol\mu})
=\,-\,{\bf b}\cdot{\bf a}\,+\,0\,=\,-\,{\bf a}\cdot{\bf b}\,
=\,-\,\cos\theta_{{\bf a}{\bf b}}\,.\label{derive-2}
\end{equation}

Well, there seems to be no difference between the results (\ref{derive-1}) and (\ref{derive-2}),
which would also be the case for commuting real functions. Thus, the above categorical assertion
made by Philippe Grangier is simply false within my model. One of the consequences of this
result, actually, is that my model is at least observationally local, just as quantum mechanics
is. In fact, despite the apparent non-commutativity of observables, my model also happens to
be {\it ontologically} local, at the level of individual microstates, as has been amply
demonstrated in Ref.\cite{further}. Thus, contrary to Grangier's claims, my model {\it does}
provide physical means to extract correct measurement results from my observables. Moreover,
in defiance of the meaningless musing in his footnote [4], my model happens to be unequivocally
local realistic.

\smallskip

Despite the above evidence, I would not be surprised if Grangier continues to shy away from
using non-commuting quaternions in a local realistic theory, and insists on averaging only
over commuting real numbers. But, as I have demonstrated in Ref.\cite{further}, nothing in the
idea of local realism---or in the reasoning of Bell---that demands that we must comply with
such an artificial restriction. A view of local realism admitting only commuting real numbers
is a very narrow and self-serving view, almost designed to protect the idea of nonlocality.
Such a view is fictional to begin with, so it is no great triumph to knock it off either by
a theorem or by an experiment. The aim of local realism is not to reject the achievements of
quantum mechanics, but to assimilate them in constructing a better map of the world.

\smallskip

Finally, Grangier expresses his unhappiness with the choice of my running title: ``Disproof of
Bell's Theorem.'' He rightly points out that as a mathematical theorem Bell's theorem cannot
be disproved, since its conclusions follow from well defined premises in a mathematically
impeccable manner. But Bell's theorem is hardly just a mathematical theorem. In fact, as a
purely mathematical theorem, it is probably one of the least interesting of all theorems. It
simply says that a certain type of commuting functions cannot reproduce a certain type of
correlations, which are known to be produced by non-commuting functions. On the other hand,
Bell's theorem, as we have come to appreciate it, is a mathematical theorem with {\it profound}
physical and metaphysical implications. To paraphrase one of the early champions of the theorem,
namely Abner Shimony, it states that: {\it no physical theory which is realistic as well as
local in a specified sense can reproduce all of the statistical predictions of quantum mechanics}
\cite{Shimony-2004}. Now this claim---with its whole package of mathematics, physics, and
metaphysics---is certainly subject to a refutation, say by means of a counterexample, no less
than the theorem of von Neumann was, until thrown on the wayside by Bohm's theory.

\bigskip

{\underline{\bf Response \# 4 --- 13 August 2007}}:

\bigskip

In response to my previous response to his critique \cite{Grangier}, Philippe Grangier has
amended his critique. Although this amendment fails to validate his arguments, it does
provide me an opportunity to further elucidate the basis of my model. Below I shall demonstrate
that his critique is based on a spurious distinction between ``commuting real functions''
and ``algebraic quantities'' on the one hand, and cosmetic similarity between the vectors
${\epsilon\,{\bf n}}$ of his ``toy model'' and the bivectors ${{\boldsymbol\mu}\cdot{\bf n}}$
of my local model on the other. As I stressed above, his confusions are a result of his
failure to understand the meaning and function of the bivectors within my model,
which are abstract quantities, quite different from the ordinary vectors such as
${\epsilon\,{\bf n}}$. Consequently, his critique misses its goal by a mile.

\smallskip

To explain these facts further, let me begin with an issue of terminology. In my previous
publications I have used the word ``observables'' to mean ``local realistic variables''. This
terminology was borrowed from the celebrated report on Bell's theorem by Clauser and Shimony
\cite{Clauser-Shimony}. Grangier finds this terminology unsatisfactory, and I concur. There is
of course Bell's favored terminology for the concept I have in mind---namely, ``local beables''
\cite{Bell-La}---but this terminology has been justifiably criticized by Shimony
\cite{Shimony-beable}, and hence I have been reluctant to use it in my papers. I will use it
here, however, to avoid the kind of misinterpretation of my model that is pervasive in
Grangier's footnotes.

\smallskip

Suppose, then, we have three non-vanishing local beables, ${A_{\bf a}(\lambda)}$,
${B_{\bf b}(\lambda)}$, and ${C_{\bf c}(\lambda)}$, with binary values. Within the local
realistic framework of Bell's theorem, these beables are traditionally taken to be commuting
real functions, providing correlation functions such as
\begin{equation}
{\cal E}_{h.v.}({\bf a},\,{\bf b})\,=\int_{\Lambda}
A_{\bf a}(\lambda)\,B_{\bf b}(\lambda)\;\,d\rho(\lambda)\,=\int_{\Lambda}
B_{\bf b}(\lambda)\,A_{\bf a}(\lambda)\;\,d\rho(\lambda).\label{prob}
\end{equation}
Now Grangier seems to think that ``commuting real functions'' are somehow distinct from
``algebraic quantities'' \cite{Grangier}. But this is simply a misconception of elementary
mathematical facts, for when we say that beables are ``commuting real functions'' or ``scalar
quantities'', what we really mean is that they satisfy the following {\it algebraic} relations:
\begin{alignat}{2}
\{-\,A_{\bf a}(\lambda)\} + A_{\bf a}(\lambda) & =
             A_{\bf a}(\lambda) + \{-\,A_{\bf a}(\lambda)\} = 0
                    && {\rm Existence\;of\;additive\;inverse\;and\;identity} \\
A^{-1}_{\bf a}(\lambda)\,A_{\bf a}(\lambda) & =
A_{\bf a}(\lambda)\,A^{-1}_{\bf a}(\lambda) = 1
                      && {\rm Existence\;of\;multiplicative\;inverse\;and\;identity} \\
\!\!\!A_{\bf a}(\lambda) + \{B_{\bf b}(\lambda) + C_{\bf c}(\lambda)\} & =
\{A_{\bf a}(\lambda) + B_{\bf b}(\lambda)\} + C_{\bf c}(\lambda)
                                      && {\rm Associative\;law\;of\;addition}\;\;\; \\
A_{\bf a}(\lambda)\,\{B_{\bf b}(\lambda)\,C_{\bf c}(\lambda)\} & =
\{A_{\bf a}(\lambda)\,B_{\bf b}(\lambda)\}\,C_{\bf c}(\lambda)
                                      && {\rm Associative\;law\;of\;multiplication} \\
A_{\bf a}(\lambda)\,\{B_{\bf b}(\lambda) + C_{\bf c}(\lambda)\} & =
A_{\bf a}(\lambda)\,B_{\bf b}(\lambda) + A_{\bf a}(\lambda)\,C_{\bf c}(\lambda)
            && {\rm Left\;distributive\;law\;of\;multiplication\;over\;addition} \\
\!\!\!\{B_{\bf b}(\lambda) + C_{\bf c}(\lambda)\}\,A_{\bf a}(\lambda) & =
B_{\bf b}(\lambda)\,A_{\bf a}(\lambda) + C_{\bf c}(\lambda)\,A_{\bf a}(\lambda) \;\;\;
            && {\rm Right\;distributive\;law\;of\;multiplication\;over\;addition} \\
A_{\bf a}(\lambda) + B_{\bf b}(\lambda) & = B_{\bf b}(\lambda) + A_{\bf a}(\lambda)
                                        && {\rm Commutative\;law\;of\;addition} \\
A_{\bf a}(\lambda)\,B_{\bf b}(\lambda) & = B_{\bf b}(\lambda)\,A_{\bf a}(\lambda)
                                        && {\rm Commutative\;law\;of\;multiplication}.
\end{alignat}

For Bell, these eight relations then define the algebra of local beables. Note that so far we
are entirely within the standard local realistic framework of Bell. I have stated the above
relations explicitly simply to emphasize the fact that the local beables of Bell are {\it no
less} algebraic than those of my model. In what follows, we shall see how one can systematically
arrive at my local beables (``bivectors'') from the above local beables of Bell (``scalar
functions''). Once this morphosis from Bell's beables to my beables is correctly
understood, and the true meaning of the word ``bivector'' is genuinely appreciated, it will
be more than clear why I continue to find Grangier's critique vacuous.

\smallskip

Now, the above algebra of the local beables of Bell happens to be a {\it normed division algebra}
over the real numbers commonly known as {\it the scalar algebra} \cite{Dixon}. Thus, when we
assert that
the beables ${A_{\bf n}(\lambda)}$ of Bell are scalar quantities or commuting real functions,
what we are asserting is that they are elements of the above scalar algebra. Indeed, a quantity
is said to be ``scalar'' {\it if and only if} ${\,}$it is an element of the above algebra.
Moreover, this algebra happens to be one of the only three possible associative division
algebras over the real numbers. There is, however, a serious deficiency in this algebra.
This deficiency is well known
to aerospace engineers and computer-vision experts, but has been entirely overlooked by the
practitioners of Bell's theorem. The deficiency is that the elements of the scalar algebra are
not capable of representing rotations in the ordinary three-dimensional space in a
{\it singularity-free} manner. When one tries to represent rotations using the elements of
scalar algebra---say by using Euler angles for example---one runs into what is known as a
{\it gimbal lock}. The problem is that when two of the three rotation axes align themselves,
then one of the rotation degrees of freedom is lost. This is a fatal problem for airplane
controls, especially if an airplane is in a steep ascent or descent. Thus, if you are a
local realist, Bell's scalar algebra of local beables is hardly the algebra you would want to
use for your theories. More importantly, if you are travelling in an airplane, you would
probably want to hope that the controls of that airplane are not designed using the algebra
of the local beables of Bell.

\smallskip

Fortunately, the above problem suggests its own solution. If one wants to represent rotations
within a local realistic theory in a singularity-free manner, then the most natural way to do
that is to relax the last of the conditions---namely, the condition (19)---of the scalar algebra
of Bell. Remarkably, when this condition of multiplicative commutativity is relaxed in a careful
manner, the resulting algebra continues to remain an associative division algebra over the real
numbers, known as the quaternionic algebra. This is of course the non-commutative algebra
famously discovered by Hamilton in the morning of the Monday, ${16^{th}}$ of October 1843. Thus,
any deterministic local beables that can model the EPR-Bohm correlations would have to be
quaternionic numbers, if they are to {\it smoothly} represent rotations in the ordinary physical
space. This is actually not surprising, since quaternions are well known to provide a spin-1/2
representation of the rotation group. But the point is that, whether one is trying to model
EPR-Bohm correlations or airplane controls, the uniqueness of the representation of rotations by
Euler angles degenerates into multivaluedness at some point (gimbal lock), while the quaternionic
representation of rotations always remains a ``double cover'', with unit quaternion and its
negative both giving the same rotation. Another way to see this is to recall
that the rotation group SO(3) is ``doubly connected'', meaning that there are loops in it which
cannot be continuously contracted to a point, whereas its ``double cover'' group, SU(2) (which
is the group of ${2\times2}$ complex unitary matrices with determinant 1), is ``simply
connected'', and is precisely the group of unit quaternions \cite{Dixon}.
Now, one usually views quaternions as a kind of generalization of the complex numbers.
But there is nothing intrinsically ``complex''
about the quaternions. Such a representation makes them unnecessarily
mysterious. In fact, within Clifford algebra ${Cl_{3,0}}$---where they are called
bivectors---quaternions emerge naturally as {\it real} quantities, and can be succinctly
defined by the product relations
\begin{equation}
(\,{\boldsymbol\mu}\cdot{\bf a})(\,{\boldsymbol\mu}\cdot{\bf b})\,
=\,-\,{\bf a}\cdot{\bf b}\,-\,{\boldsymbol\mu}\cdot({\bf a}\times{\bf b}).\label{bibi-identity}
\end{equation}
Here ${\boldsymbol\mu}$ is a unit trivector with an unspecified orientation. The reason why the
handedness of the trivector ${\boldsymbol\mu}$ has not been fixed here by setting it equal to
${+\,I}$ or ${-\,I}$ is because such a choice is not an intrinsic feature of the algebra
${Cl_{3,0}}$. In text books the handedness of ${\boldsymbol\mu}$ is always extraneously imposed
on the algebra, as an {\it ad hoc} choice. Now, what the above condition tells us is that, if one
wants to represent rotations in a {\it singularity-free} manner, then a local realistic way to
do this is to replace the commutativity condition (19) of Bell's algebra of local beables with
the condition
\begin{equation}
[\,A_{\bf a}(\lambda),\,B_{\bf b}(\lambda)\,]\,
=\,-\,2\,C_{{\bf a}\times{\bf b}}(\lambda)\,,\label{bibi-replace}
\end{equation}
making the resulting algebra a bivector subalgebra of the Clifford algebra ${Cl_{3,0}}$. What
is more, this resulting algebra continues to satisfy the remaining conditions---the conditions
(12) to (18)---of the scalar algebra of Bell, and is nothing but the quaternionic algebra in a
local realistic form. In other words, contrary to Grangier's claim, Bell's beables do not cease
to be local realistic just because the condition (19) of the scalar algebra is replaced by a
more satisfactory condition (\ref{bibi-replace}). The fact that my beables defined by the
bivector subalgebra of the algebra ${Cl_{3,0}}$ are strictly local has been rigorously proved
in Ref.\cite{further}. And the fact that they remain realistic can be checked by rotating a book.
When I rotate a book first about the x-axis and then about the y-axis by 90 degrees, the net
result is different from the case when I restore its initial state and rotate it again about
the two axes in the reverse order. But this does not, of course, render either the book,
the rotational states of the book, or the processes of its rotation any less realistic.

\smallskip

Let me retrace my steps so far to make sure I have addressed all the issues raised by Grangier.
I began with the algebra of the local beables of Bell, and noticed that his scalar algebra is not
capable of representing rotations in a singularity-free manner in the three-dimensional physical
space. This observation led me to relax the last condition of Bell's scalar algebra, thus
transforming his algebra into the algebra of quaternions. Since I am interested in a local
realistic model, the correct representation of this algebra is the bivector subalgebra of
the Clifford algebra ${Cl_{3,0}}$. The resulting beables are then bivectors, which are both
{\it local} and {\it realistic}. This, then, makes it quite clear where the local beables of my
model came from. In particular, this makes it clear that the bivectors
${{\boldsymbol\mu}\cdot{\bf n}}$ within my model---which Grangier pejoratively calls ``algebraic
quantities''---are {\it not} all that different from the commuting local beables of Bell. In
fact, apart from the multiplicative
non-commutativity, they satisfy exactly the same algebra as do the local beables of
Bell. What is more, far from being {\it ad hoc}---as Grangier would have us believe---the
local beables of my model result directly from a {\it necessary} correction to the algebra of
the local beables of Bell. They are necessary to represent rotations in the physical space in a
singularity-free manner. Thus, there is not a single step which can be said to be
``{\it ad hoc}'' between my starting point---namely, the algebra of local beables of Bell
defined by the conditions
(12) to (19)---and my end point---namely, the local beables of my model defined by the following
corrected algebra:
\begin{alignat}{2}
\{-\,A_{\bf a}(\lambda)\} + A_{\bf a}(\lambda) & =
             A_{\bf a}(\lambda) + \{-\,A_{\bf a}(\lambda)\} = 0
                    && {\rm Existence\;of\;additive\;inverse\;and\;identity} \\
A^{-1}_{\bf a}(\lambda)\,A_{\bf a}(\lambda) & =
A_{\bf a}(\lambda)\,A^{-1}_{\bf a}(\lambda) = 1
                      && {\rm Existence\;of\;multiplicative\;inverse\;and\;identity} \\
\!\!\!A_{\bf a}(\lambda) + \{B_{\bf b}(\lambda) + C_{\bf c}(\lambda)\} & =
\{A_{\bf a}(\lambda) + B_{\bf b}(\lambda)\} + C_{\bf c}(\lambda)
                                      && {\rm Associative\;law\;of\;addition}\;\;\; \\
A_{\bf a}(\lambda)\,\{B_{\bf b}(\lambda)\,C_{\bf c}(\lambda)\} & =
\{A_{\bf a}(\lambda)\,B_{\bf b}(\lambda)\}\,C_{\bf c}(\lambda)
                                      && {\rm Associative\;law\;of\;multiplication} \\
A_{\bf a}(\lambda)\,\{B_{\bf b}(\lambda) + C_{\bf c}(\lambda)\} & =
A_{\bf a}(\lambda)\,B_{\bf b}(\lambda) + A_{\bf a}(\lambda)\,C_{\bf c}(\lambda)
            && {\rm Left\;distributive\;law\;of\;multiplication\;over\;addition} \\
\!\!\!\{B_{\bf b}(\lambda) + C_{\bf c}(\lambda)\}\,A_{\bf a}(\lambda) & =
B_{\bf b}(\lambda)\,A_{\bf a}(\lambda) + C_{\bf c}(\lambda)\,A_{\bf a}(\lambda) \;\;\;
            && {\rm Right\;distributive\;law\;of\;multiplication\;over\;addition} \\
A_{\bf a}(\lambda) + B_{\bf b}(\lambda) & = B_{\bf b}(\lambda) + A_{\bf a}(\lambda)
                                        && {\rm Commutative\;law\;of\;addition} \\
[\,A_{\bf a}(\lambda),\,B_{\bf b}(\lambda)\,] & =\,-\,2\,C_{{\bf a}\times{\bf b}}(\lambda)
                                     && {\rm Non\!\!-\!\!commutative\;law\;of\;multiplication}.
\end{alignat}
What is more, this corrected algebra of beables happens to be quite {\it unique}, in the
sense that it is the only possible non-commutative algebra that remains an associative normed
division algebra over the field of real numbers. Bell's algebra, by contrast, is a
{\it commutative} and associative normed division algebra over the field of real numbers.

\smallskip

The fact that my local beables---namely the bivectors ${{\boldsymbol\mu}\cdot{\bf n}}$---are
anything but {\it ad hoc}, and the fact that apart from the non-commutativity of their
products they are no different from Bell's local beables, can be further elucidated in the
light of the ``toy model'' introduced by Grangier. The basic quantities in this toy model are
the ordinary vectors ${\epsilon\,{\bf n}}$, apart from the sign ambiguity ${\epsilon=\pm\,1}$.
Grangier compares this {\it ad hoc} quantities ${\epsilon\,{\bf n}}$ with my carefully thought
out local beables ${{\boldsymbol\mu}\cdot{\bf n}}$, in an attempt to evade the inescapable
conclusions of my preprints. But the comparison of two quantities can hardly be
more misguided. To be sure, there do exist {\it superficial} ${\,}$similarities between
${\epsilon\,{\bf n}}$ and ${{\boldsymbol\mu}\cdot{\bf n}}$. There is of course the sign
ambiguity in both quantities, and there appears to be an ordinary vector ${\bf n}$ involved in
both of them. But the similarities end there. One must not judge a book by its cover.
Apart from the sign ambiguity, the toy quantity ${\epsilon\,{\bf n}}$ is simply an ordinary
vector, whereas the local beable ${{\boldsymbol\mu}\cdot{\bf n}}$ is an entirely different
object.

\smallskip

To begin with, the sense of rotation quantified by the signs ${+}$ or ${-}$ is an {\it
intrinsic} feature of any bivector, unlike the case in Grangier's {\it ad hoc} toy quantity
${\epsilon\,{\bf n}}$. More significantly, despite the appearance, the vector ${\bf n}$
is {\it not} a part of the bivector ${{\boldsymbol\mu}\cdot{\bf n}}$ at all. In other words,
despite its appearance in ${{\boldsymbol\mu}\cdot{\bf n}}$, the vector ${\bf n}$ is actually
{\it not there}. In the case of the quantity ${\epsilon\,{\bf n}}$ on the other hand, the
vector ${\bf n}$ is almost all what is there \cite{Cliff}. One must not be misled by the notation
${{\boldsymbol\mu}\cdot{\bf n}\equiv{\boldsymbol\mu}\,{\bf n}}$, which simply indicates the
{\it duality} relation between the vector ${\bf n}$ on the one hand---indicating the direction
{\it perpendicular} to the bivector---and the bivector ${{\boldsymbol\mu}\cdot{\bf n}}$ itself
on the other. In fact, the bivector ${{\boldsymbol\mu}\cdot{\bf n}}$ itself represents nothing
more than {\it abstractions} of a planar orientation and magnitude, together with an intrinsic
sense of rotation. In particular, the shape of its plane segment is of no significance. What
is more, a bivector denoted by the wedge product
${{\bf a}\wedge{\bf b}}$ is fundamentally different from an axial vector denoted by the
cross product ${{\bf a}\times{\bf b}}$. The former is {\it intrinsic} to the plane containing
${\bf a}$ and ${\bf b}$, and remains {\it independent} of the dimension of a vector space in
which it lies. Consequently, comparing an ordinary 3-vector such as ${\epsilon\,{\bf n}}$ with
the bivector ${{\boldsymbol\mu}\cdot{\bf n}}$ is almost a {\it non sequitur}.
A bivector itself, as an {\it abstract} quantity, is characterized by three properties and three
properties {\it only}. These are: (1) its magnitude, (2) its orientation, and (3) its sense of
rotation. Now, the magnitude of the bivectors ${{\boldsymbol\mu}\cdot{\bf n}}$ of my local model
is universally fixed to be unity. Their orientations, on the other hand, are defined by the
directions ${\bf n}$ of the spin analyzers themselves. But the purpose of this dual direction
${\bf n}$ is to simply provide a useful intuitive picture for the alignment with a spin analyzer,
and it should not be confused with any intrinsic property of the bivector itself. Once any
such confusion is exorcised, it becomes obvious that all that finally remains of the bivector
itself is its intrinsic sense of rotation {\it about} the direction {\bf n}, indicated by the
${+}$ or ${-}$ signs, corresponding to the observed ``measurement results.'' So, forgive me for
losing my collegial tone, but from the perspective of my model the whole business of ``a way to
extract the ``sign'''' that Grangier is so concerned about is profoundly misguided. It stems
from a confusion between an ordinary vectorial quantity such as ${\epsilon\,{\bf n}}$ and the
sophisticated notion of a bivector. In other words, given the geometrical necessities discussed
above on how rotations in the physical space must be modeled, and from the corresponding
perspective of the algebra ${Cl_{3,0}}$, the RHS of the displayed expression in his footnote [7]
is {\it absurd}.

\smallskip

From the above discussion it is clear that Grangier's entire critique of my model is based
on multiple misconceptions about the basic concepts involved in my model. It is then not
surprising why he has been led astray in his interpretation of my model. Consider,
for example, the assertions in his footnote [4]. Now there are several problems with this
footnote. To begin with, I am not happy with the way a partial quotation from one of my preprints
appears in it out of context. But respecting the contextuality of text does not seem to be
one of Grangier's fortes, so perhaps I should let this problem slide. The main claim in this
footnote seems to be that my model is ``...``non-realistic'', since one cannot associate
simultaneous ``elements of reality'' to the two non-commuting measurements ${a}$ and ${a'}$
on one side (or ${b}$ and ${b'}$ on the other side).'' Since I am unable to relate this sentence
to the words from my text that have been used to support it, I will presume that what
Grangier has in mind here is the non-commutativity of my beables given by
\begin{equation}
[\,{\boldsymbol\mu}\cdot{\bf a},\,{\boldsymbol\mu}\cdot{\bf a}'\,]\,
=\,-\,2\,{\boldsymbol\mu}\cdot({\bf a}\times{\bf a}')\,,\label{non-identity}
\end{equation}
which indeed holds within my model, but {\it only counterfactually} \cite{further}. Moreover,
this {\it counterfactual non-commutativity} has nothing whatsoever to do with the quantum
mechanical indeterminacies. It simply reflects the well known geometrical fact that
compositions of rotations in ordinary physical space do not commute in general. If I rotate
a book about the direction ${\bf a}$ by a certain amount, and then about the direction
${{\bf a}'}$ by the same amount, I get one result. But if I start all over again and perform
the same two rotations in the reverse order, I get a different result. There is no mystery
here. There is no question of not having ``elements of reality'' in whatever sense, because
equation (\ref{non-identity}) is simply a geometrical constraint that tells us which elements
of reality are consistent with the geometry of the physical space. If one nevertheless imagines
some {\it a priori} ${\,}$elements of reality inconsistent with the geometry of the physical
space, then those ``elements of reality'' are fictitious to begin with, and hence they should
not be of any concern to the local realist. In other words, as we discussed above, the very
purpose of having the condition (\ref{non-identity}) within the algebra of local beables is to
make sure that {\it correct} elements of reality are assigned to the rotational degrees of
freedom, in a manner {\it more consistent} than can be done within the commutative algebra of
the local beables of Bell.

\smallskip

The same theme continues in Grangier's footnote [8], where he states: ``...knowing the ``sign''
of ${(\mu\cdot a)}$ forbids to tell anything about the ``sign'' of ${(\mu\cdot a')}$, for the
same given ${\mu}$...'' But this assertion is simply false, because, again, the
non-commutativity of local beables happens to hold only counterfactually within my model.
In particular, if one can manage it in practice, nothing within the model itself
forbids simultaneous measurements of the beables ${{\boldsymbol\mu}\cdot{\bf a}}$ and
${{\boldsymbol\mu}\cdot{\bf a}'}$, since in that case, as discussed in Ref.\cite{further},
the RHS of equation (\ref{non-identity}) would vanish identically (and hence my local beables
would essentially reduce to those of Bell, satisfying the conditions (12) to (19)). I suppose,
however, that Grangier would want to assume truth-values even for counterfactual conditionals.
But in that case, even when the RHS of equation (\ref{non-identity}) happens to be non-zero,
its function is to simply encode how rotations are composed within the physical space. Thus,
even if one views the condition (\ref{non-identity}) as a kind of constraint on counterfactual
realizability of the values of both local beables (whatever that means), it is not a constraint
coming from nowhere. It is simply a constraint imposed by the geometry of the physical space
itself, dictating how rotations in the physical space ought to be composed. To conclude from
this that my model is somehow ``non-realistic'' is as absurd as concluding that the orthogonal
directions in the physical space are ``non-realistic'', simply because their products do not
commute.

\smallskip

In conclusion, the catalogue of confusions and misinterpretations in Grangier's critique of my
model suggests a failure on his part to correctly understand the notion and function of
bivectors within the Clifford algebraic framework. Witness for example his footnote [5], where
he seems to think that the vector ${\bf a}$ is somehow an intrinsic part of the bivector
${{\boldsymbol\mu}\cdot{\bf a}}$ rather than its auxiliary assistant. In reality, however,
a bivector is a sophisticatedly defined abstract entity,
whose function is to correctly encode how rotations are composed within our physical space on
the one hand, and precisely predict the binary results of spin measurements on the other.
Far from being a mysterious ``algebraic quantity'', it naturally emerges as an element of a
{\it corrected} algebra of the local beables of Bell. Far from being introduced in an {\it ad
hoc} ${\,}$manner, it has been systematically arrived at, starting from the algebra of the local
beables of Bell himself. Therefore, it is entirely inappropriate to compare it with a vectorial
toy quantity such as ${\epsilon\,{\bf a}}$.

\bigskip

{\underline{\bf Response \# 5 --- 19 August 2007}}:

\bigskip

In the third version of his critique, Grangier has revised his conclusion \cite{Grangier}.
Despite the revised conclusion however, I continue to reject his critique, for it continues to
miss the forest for the trees. Notwithstanding the importance of the second-level questions
he has raised, what is truly at stake here is a question of principle. In my view, the real
significance of Bell's theorem lies in its far-reaching physical and metaphysical
conclusions---namely, that {\it no local realistic theory can reproduce all of the statistical
predictions of quantum mechanics}. One would be hard-pressed to find an impossibility claim as
far-reaching as this in the entire history of physics. But it is precisely the scope of this
claim that makes Bell's theorem vulnerable.
For, surely, it can be refuted if one can construct a local realistic model that reproduces
all of the predictions of quantum mechanics for the relevant physical scenario. More precisely,
the crucial question concerning the theorem is whether or not it is possible to construct a
local realistic model for the EPR-Bohm correlations that can
reproduce all of the relevant predictions of quantum mechanics. In contradiction to the
theorem, an affirmative answer to this question has been provided in
Refs.\cite{Christian} and \cite{further}. In particular, it has been demonstrated in these
references that a deterministic, local realistic model exists, which {\it exactly} reproduces
{\it all} of the predictions of quantum mechanics relevant for the EPR-Bohm experiments,
including correlations and binary results (${\pm\,1}$) of spin measurements. {\it If}
${\,}$Grangier's toy quantity ${\epsilon\,{\bf n}}$ can also be shown to accomplish {\it all} of
the same results, then I would certainly welcome it as a second disproof of Bell's theorem. And
from what I gather from my readings of Bell's own views, he would welcome it too: ``...what is
proved by impossibility proofs is lack of imagination \cite{Bellimpos}.''

\vfill\eject

\bigskip

{\underline{\bf Response \# 6 --- 12 December 2007}}:

\bigskip

Tung Ten Yong \cite{Tung} seems to have deemed it appropriate to allow a mathematical howler
within his own critique, if that is what it takes to allege an error in my argument.
The focus of his complaint is my expression 
\begin{equation}
\int_{{\cal V}_3}(\,{\boldsymbol\mu}\cdot{\bf a}\,)
(\,{\boldsymbol\mu}\cdot{\bf b}\,)\;\,d{\boldsymbol\rho}({\boldsymbol\mu})\,,\label{derive-5}
\end{equation}
which is an average over local beables ${{\boldsymbol\mu}\cdot{\bf a}}$ and
${{\boldsymbol\mu}\cdot{\bf b}}$ with binary outcomes, analogous to the expectation value of such
beables considered by Bell. Now Tung Ten Yong claims that, since the integrand in the above
expression is a multivector quantity rather than a scalar quantity, the result of the integration
would not in general be an expression in accordance with the orthodox understanding of what such an
expression is supposed to be. However, apart from failing to understand the {\it modus operandi}
of a counterexample, what he fails to realize is that the measure
${d{\boldsymbol\rho}({\boldsymbol\mu})}$ in this
expression---which he writes as a scalar ${d{\rho}({\boldsymbol\mu})}$ exposing the extent of his
ignorance---is no ordinary measure. It is, in fact, a {\it directed} measure. That is to say,
${d{\boldsymbol\rho}({\boldsymbol\mu})}$ itself is not a scalar but a multivector.
Moreover, the juxtaposition of the integrand and its measure in the above expression is
no ordinary product, but a {\it Clifford} product, between the multivectors
${({\boldsymbol\mu}\cdot{\bf a})({\boldsymbol\mu}\cdot{\bf b})}$ and
${d{\boldsymbol\rho}({\boldsymbol\mu})}$. In other words, the above expression is a
{\it directed} integral of multivectors, with the integrand and measure both being multivector
quantities, producing the desired scalar quantity upon average. If instead, as in Tung Ten Yong's
Eq.(1), one na\"ively attempts to integrate a multivector quantity with a {\it scalar} measure, then
nonsense surely results. On the other hand, when consistently employed, the Clifford-algebraic notion
of directed measure is far more empowering, and easily permits generalization to the cases where the
distribution ${{\boldsymbol\rho}({\boldsymbol\mu})}$ is non-isotropic. An explicit example of such a
generalization can be found in Sec.V of Ref.\cite{further}. Now directed measures and directed integrals
are rather subtle concepts within Clifford algebras, requiring more than just a casual acquaintance with
the subject. If one is not familiar with these concepts, then it is prudent to first learn about them
from a good textbook \cite{Sobczyk}, before making a sophomoric howler (as in \cite{Tung}). In particular,
as is evident from the explicit example cited above, it is not the integrand by itself, or its directed
measure, but the Clifford product of the two as a whole that is responsible for achieving the scalar
quantities in general. That mathematically such scalar quantities can always be achieved by appropriately
choosing local beables and directed measures is beyond doubt, thanks to the invertibility of the
Clifford product \cite{Sobczyk}. That it can also be achieved in a physically meaningful manner for the
rotationally invariant entangled state \cite{Christian} (as well as for the Malus's law, despite
anisotropy \cite{further}) is more than sufficient for the purposes of producing
a counterexample to Bell's theorem. Whereas the question whether this can also be achieved for
{\it all} conceivable physical scenarios {\it beyond} Bell's theorem, is beyond the purposes of my papers.

\smallskip

The remaining of the discussion in Tung Ten Yong's critique, and especially his Eq.(3), reveals that,
despite my repeated explanations in the past (cf. \cite{further}), he has yet to understand the nature
and function of my local beables.

\smallskip

Finally, it is worth noting here that in the revised versions of his preprint Tung Ten Yong has been driven
to commit further (even more na\"ive) blunders than the ones revealed above. For instance, contrary to
his absurd assertions, the net result ${{\bf n}-{\bf n}=0}$ in equation (18) of my preprint \cite{Christian}
is not some non-trivial null vector, but a {\it zero} vector, with {\it all} of its components being
{\it strictly zero}. As is well known, non-zero null vectors do not exist in the Euclidean space. It
would be extraordinary if the fate of local realism were to hinge on a technical distinction
between vanishing scalars and strictly zero vectors. It would be as
if the fate of Newton's theory of mechanics were to hinge on the equation ${{\bf F}-{\bf F}=0}$
being interpreted, not as a zero force, but as no force at all. Fortunately it will not come to that,
since scalars are as much a part of the Clifford algebra as any other multivectors (they are duals of
trivectors and span a one-dimensional subspace of the space ${Cl_{3,0}}$ \cite{Cliff}). In fact, in
mockery of the assertions by Tung Ten Yong, this inclusion of scalars among the subspaces of
${Cl_{3,0}}$ diminishes the artificial distinction between scalars in the field ${\rm I\!R}$
and vectors in the space ${{\rm I\!R}^n}$ that
is usually found in the definition of a vector space \cite{Dorst}. Indeed, the most general form of a
multivector in ${Cl_{3,0}}$ includes a scalar explicitly---and {\it essentially}---in defiance of the
strict segregation of scalars and vectors maintained in the vector algebra. In summary, the merits of
a Clifford-algebraic model ought to be judged from the Clifford-algebraic perspective, not from some
misconceptions one may happen to harbor.

\smallskip

{\underbar{\bf Note added in response to version 4}:} As far as I can see, I have more than adequately
addressed Tung Ten Yong's claim of a so-called error in my paper; but since my response has not been
understood, let me spell it out once again in simpler terms. The task of a local realist seeking a
counterexample to Bell's theorem is to choose his or her local beables in such a manner that they
go hand-in-hand with the chosen distribution of hidden variables and its measure. If this is done
carefully, then the relevant predictions of quantum mechanics can be reproduced in a local realistic
manner. If, however, the chosen distribution happens to be non-isotropic, then the corresponding local
beables are unlikely to be those chosen to match the isotropic distribution. This much is elementary,
and holds true whether or not the local beables are chosen to be Clifford-algebraic. On the other
hand, if the local beables {\it are} chosen to be Clifford-algebraic, then it is even more imperative
that they are correctly matched with the distribution of hidden variables and its measure, because
otherwise not only would it be impossible to reproduce the predictions of quantum mechanics, but also the
whole construction would be incongruous with the basic framework of Clifford algebras. Moreover, once the
Clifford-algebraic framework is chosen for the task, the integrity of the framework must be maintained,
both mathematically and conceptually. In particular, it makes no sense within this framework to
speak of a ``scalar codomain'' for the integrand (local beables) alone, as was done in the preprint
\cite{Tung} before prompted for corrections. With these remarks, the flaws in the reasoning of Tung Ten
Yong are now easy to spot. He alleges a mathematical inconsistency in my paper by considering a hypothetical
non-isotropic distribution, but at the same time using those local beables that have been chosen to match an
isotropic distribution. This is quite an odd sort of reasoning, for at the end of the day the goal of a local
realist is not to produce a general non-zero multivector, but a zero multivector. If the chosen local beables
are not up to the job, then one throws them away and looks for another set. It is absurd to claim that, since
the local beables chosen for the isotropic distribution do not give the right answer for the non-isotropic
distribution, the whole scheme is somehow inconsistent. If the point being made by this odd reasoning is to
show that the integrand and its measure in the expression (\ref{derive-5}) above are multivector quantities,
then that has never been a secret in my papers. Now, from the perspective of the orthodox ideas that keep the
notions of scalars and vectors strictly separate from each other---such as the ideas Tung Ten Yong insists on
employing---the choice of multivectors as local beables and measures would certainly look unorthodox. As discussed
above, however, such an orthodox distinction between scalars and other multivectors is dramatically blurred within
Clifford algebras, and this blurring is anything but an ``error''. In fact, it is absolutely essential for the
completeness of the resulting Clifford-algebraic structure. More importantly, a fully rigorous framework 
of geometric calculus exists \cite{Sobczyk}, precisely to deal with the directed integrals such as (\ref{derive-5})
where the integrand and measure are both multivector quantities. And it is this rigorous framework which is consistently
used to evaluate the integrals (18) and (19) of my preprint \cite{Christian}. No other concepts extraneous to the
Clifford-algebraic framework are necessary for this task. Now, after several versions of his
preprint, what seems to remain of the ``objection'' of Tung Ten Yong is the trivial fact that, even for the right
combination of the integrand and its isotropic measure, what results after the integration in Eq.(18) of
\cite{Christian} is a zero ``vector, not a zero scalar.'' But within Clifford algebras this alleged distinction
between ``zero scalar'' and ``zero vector'' has no meaning. Indeed, the word ``scalar'' within this framework
simply means a multivector whose all but one components are strictly vanishing. In other words, a ``scalar''
within the algebras ${Cl_{3,0}}$ is a multivector whose vector, bivector, and trivector components are all
strictly zero. Thus, the end result of the integration in Eq.(18) of preprint \cite{Christian} is nothing but the
Clifford-algebraic counterpart of the vector-algebraic notion of a ``zero scalar.'' In other words, from the
Clifford-algebraic perspective, the entire discussion by Tung Ten Yong is rather frivolous.

\smallskip

A final point needs to be made about some further blunders and misstatements in the version 4 of the preprint. The
most important of this concerns the erroneous reading of the directed measure ${d{\boldsymbol\rho}({\boldsymbol\mu})}$
and the Eq.(19) of my preprint \cite{Christian}. Tung Ten Yong alleges that the third line of this equation does not
result in the Clifford-algebraic scalar
${-{\bf a}\cdot{\bf b} + 0}$ as written in my preprint, but instead results in a trivector. Once again, the
error he is making in this claim is so obvious and so trivial that I will not insult his intelligence by
pointing it out to him. I will let him discover it for himself, and then withdraw his comments. Since this dialogue
is turning into an education project in Clifford algebra and geometric calculus in which I do not wish to
participate, this will be my last response to his preprint.

\smallskip

{\underbar{\bf Note added in response to version 5}:} Since Tung Ten Yong continues to accuse me of skirting his
``objections'' and not recanting the ``errors'' in my papers, I will try one last time to make him realize that the
objections he has raised are, in fact, nonsensical. They are a result of his own uninformed and erroneous adaptation
of the probability theory to the Clifford-algebraic framework employed in my papers. Once I spell out how an adequate
formulation of the probability theory has been correctly applied within my papers, the na\"ivety of his claims will
become evident.

\smallskip

To this end, let me begin by recalling that the focus of my papers is on the very first theorem by John Bell, which
concerns the deterministic local hidden variable theories. In other words, the focus of my papers is on the expectation 
value of a pair of deterministic local beables ${A_{\bf a}(\lambda)}$ and ${B_{\bf b}(\lambda)}$, given by the standard
probabilistic expression
\begin{equation}
{\cal E}_{h.v.}({\bf a},\,{\bf b})\,=\int_{\Lambda}
A_{\bf a}(\lambda)\,B_{\bf b}(\lambda)\;\,d\rho(\lambda).\label{prob-for}
\end{equation}
Now, as von Neumann realized long ago in his pioneering analysis of hidden variables \cite{von}, the most natural and
cogent interpretation of this expectation is in terms of a linear functional on the set of {\it general}--valued
random functions. He observed that no matter which model of physics one is concerned with---the quantum mechanical
model, the hidden variables model, or any other---for theoretical purposes all one needs to consider are the expectation
values of the observables measured in various states of the system. Furthermore, as is well known since his pioneering
work, not only can one reformulate the classical probability theory in terms of expectation functionals without
recourse to the Kolmogorov formalism, but that such a reformulation is {\it equivalent} to the latter, as has been
established rigorously by Segal \cite{Segal}. In other words, the necessary mathematical foundations towards an adequate
formulation of the classical probability theory for the purposes of my paper were already laid out long ago by von
Neumann and Segal. 

\smallskip

Since Tung Ten Yong seems to be oblivious to this basic background knowledge in foundations of physics, let me spell out 
some of the details of the von Neumann-Segal formulation of the probability theory. In their conception of the theory,
instead of a probability space one begins with a strongly continuous linear functional,
\begin{equation}
{\cal E}:\,A\longmapsto {\cal E}(A),\label{linearfunc}
\end{equation}
defined on the set of general-valued functions such as ${A(\lambda)}$ [with ${\Lambda\ni\lambda}$ being the measure
space], such that
\begin{description}
 \item[\;\;\;\;\;\;\;(a)\;\;] ${{\cal E}(e)=1}$ for the unit element ${e}$ in the set of ${A}$'s,
 \item[\;\;\;\;\;\;\;(b)\;\;] ${{\cal E}(A^*A)\geq 0\;\,\forall\;A\,}$ with ${{\cal E}(A^2)=0}$ only if ${A=0}$,
 \item[{\rm and}\;\;(c)\;\;] ${{\cal E}(B^*A\,B)\leq \alpha_{{\!\!}_{A}}{\cal E}(B^*B)}$ for some real constant
${\alpha_{{\!\!}_{A}}}$.
\end{description}
Here the ``${\,*\,}$'' represents an appropriate conjugation operation (such as the ``reverse'' operation
``${\,\dagger\,}$'' in the case of the Clifford-algebra-valued functions). Now there are several important points
worth stressing here. To begin with, in this formulation of the probability theory the codomain of the functions
${A(\lambda)}$ is {\it not} restricted to be just ${\rm I\!R}$. Indeed, the very purpose of this more flexible version
of the theory is to accommodate all {\it general}--valued probability functions. In fact, in this formulation of the
theory there are no mathematical restrictions even on the codomain of the functional ${\cal E}$ itself, apart from the
defining conditions specified above. That is to say, from the purely mathematical point of view even the codomain of the
above functional ${\cal E}$ does not have to be ${{\rm I\!R}}$. These facts of course render the entire argument by Tung
Ten Yong quite vacuous from the start. What is more, as a consequence of the above noted flexibilities, the domains of
the admissible random functions ${A(\lambda)}$ in this formulation turn out to be very diverse indeed. In particular,
these functions do not have to be the usual real-valued functions of the kind Tung Ten Yong has found in the
elementary textbooks on probability theory. They can be any general-valued functions whatsoever, such as---for
example---the bounded
operators on a Hilbert space in the case of quantum mechanics, or the Clifford-algebra-valued functions in the case of
my model. And of course they can also be the plain old commuting functions considered by Bell. Indeed, the most
familiar application of this formulation of the probability theory is in the Gelfand-Na\v{\i}mark-Segal
(GNS) construction within algebraic quantum field theory, where probability functions are taken to be the abstract
${C^*}$-algebra-valued functions \cite{Murphy}. These observations further lay bare the na\"ivety of Tung Ten Yong's
reasoning, who seems to believe that only the ``usual numbered probability functions'' have a proper place in the
probability theory. Finally, it is also worth stressing here that there are no prior restrictions on the measure used in
the above expectation functional either, beyond the usual requirements placed on any Lebesgue measure. If, however, the
general-valued functions ${A(\lambda)}$ and ${B(\lambda)}$ happen to be non-commuting (as in the case of the
Clifford-algebra-valued variables of my model), then the requirement of commutativity is promoted to the level of the
expectation functional itself,
\begin{description}
 \item[\;\;\;\;\;\;\;\;\;(d)\;\;] ${{\cal E}(AB)={\cal E}(BA)}$, whether or not ${[A,\,B]=0}$,
\end{description}
and the usual probabilistic concepts and results follow through without difficulty \cite{Segal}. The fact that this 
last condition also holds within my model is seen readily from the Eqs.${\,}$(\ref{derive-1}) and (\ref{derive-2}) above.
On the other hand, the condition is not necessarily satisfied within quantum mechanics, but even in that case the above
formulation of the classical probability theory can be rigorously extended to a non-commutative version of the theory
without difficulty \cite{noncomt}.

\smallskip

Since the above background is common knowledge among the students in foundations of physics, I did not bother to
review it in my papers. I would have opted otherwise however, had I known that someone completely ignorant of this
background would challenge my papers relying solely on the textbook versions of the probability theory. In any case,  
it is within this background that Bell famously formulated his first theorem, and it is within this very same
background that the arguments in my papers have also been formulated. It should be clear by now, however, that
in the case of my model it is all the more imperative that one follows this flexible formulation of the probability
theory, rather than the rigid textbook version Tung Ten Yong is na\"ively trying to impose upon my model. Indeed,
in the case of the Clifford-algebra-valued variables of my model, the above expectation functional ${\cal E}$
takes the natural form,
\begin{equation}
{\cal E}:\,Cl_{3,0}\longrightarrow {\mathbb{S}},\label{linearcliff}
\end{equation}
where ${Cl_{3,0}}$ is the eight-dimensional Clifford space of the general multivector functions such as
\begin{equation}
{\boldsymbol\xi}\,=\,\xi_1\,+\,\xi_2\;{\bf e}_x\,+\,\xi_3\;{\bf e}_y\,+\,\xi_4\;{\bf e}_z
\,+\,\xi_5\;{\bf e}_y\wedge{\bf e}_z\,+\,\xi_6\;{\bf e}_z\wedge{\bf e}_x\,
+\,\xi_7\;{\bf e}_x\wedge{\bf e}_y+\,\xi_8\;{\bf e}_x\wedge{\bf e}_y\wedge{\bf e}_z\,,\label{eight}
\end{equation}
along with the codomain
\begin{equation}
{\mathbb{S}}\,=\,\left\{\,\xi_k\,|\;\xi_1\in[-1,\,+1]\subset{\rm I\!R}, \,{\rm with}\;\,
\xi_{k\not=1}\equiv0\,\right\}.\label{codomaincliff}
\end{equation}
It is evident from these expressions that, although there are no prior restrictions on the possible codomain of ${\cal E}$
within this formulation of the probability theory, in my model it nevertheless {\it is} effectively equal to the desired
subset of ${\rm I\!R}$:
\begin{equation}
{\mathbb{S}}\,=\,[-1,\,+1]\subset{\rm I\!R}\,.\label{codsuch}
\end{equation}
And this is quite consistent with the fact that a ``scalar'' in ${Cl_{3,0}}$ simply means a multivector whose
all but one components are strictly zero: ${\xi_{k\not=1}\equiv0}$.
So much for the ``error'' Tung Ten Yong claims to have found in my papers.

\smallskip

Let me now turn to his own elementary error, which I had hoped he would discover himself after my previous response,
but since he has failed to do so, I have no choice but to spell it out for him. The issue concerns the concept of a
directed measure ${d{\boldsymbol\rho}({\boldsymbol\mu})}$ used in my paper. A directed measure associates a direction
and a dimension as well as a magnitude to a set, and is a generalization of the scalar measure
${|d{\boldsymbol\rho}({\boldsymbol\mu})|}$, with
${d{\boldsymbol\rho}({\boldsymbol\mu})\,=\,I\,|d{\boldsymbol\rho}({\boldsymbol\mu})|}$. In my
paper I have assumed that the probability distribution ${{\boldsymbol\rho}({\boldsymbol\mu})}$ is normalized on the
manifold ${{\cal V}_3}$, providing the condition
\begin{equation}
\int_{{\cal V}_3}\;d{\boldsymbol\rho}({\boldsymbol\mu})\,=\,1.\label{norm}
\end{equation}
Tung Ten Yong claims that this assumption is unjustified, since a summation of trivectors should give a trivector, not
a scalar, and it is my mistake to take the third line of Eq.(19) of my preprint \cite{Christian} to be a scalar rather
than a trivector. Sadly, however, this observation of his is just as superficial and careless as all of his other
observations. Once again it serves to expose
the extent of his ignorance and na\"ivety. An attentive reader would have noticed at this point that the manifold
${{\cal V}_3}$ in question is in fact a {\it vector} manifold---i.e., a manifold whose {\it points} ${\,}$are
{\it vectors} in the Euclidean space ${\mathbb{E}_3}$ \cite{Sobczyk}. Consequently, with ${I^2=-1}$ and
${d{\boldsymbol\rho}({\boldsymbol\mu})\,=\,I\,|d{\boldsymbol\rho}({\boldsymbol\mu})|}$ in mind, the above normalization
condition simply amounts to the observation that the net integration with the scalar
${|d{\boldsymbol\rho}({\boldsymbol\mu})|}$ over this vector manifold is given by
\begin{equation}
\int_{{\cal V}_3}\;|d{\boldsymbol\rho}({\boldsymbol\mu})|\,=\,-\,I,\label{normscal}
\end{equation}
which expresses the fact that the total volume of the manifold ${{\cal V}_3}$ is a left-handedly-directed unit volume,
${-\,I}$.
By the way, the sign of the handedness of this directed volume is merely conventional, and can be chosen to be positive
if so wished, by simply choosing a left-handed trivector (instead of the right-handed one chosen in my paper) to perform
integrations. If this hint is still not sufficient for Tung Ten Yong to recognize his error, then nothing will.

\smallskip

It should be abundantly clear by now that there are {\it no} mathematical inconsistencies in the manner
I have employed the probability theory within my papers. That is to say, contrary to the uneducated claims made by
Tung Ten Yong, no stricture of the probability theory has been unduly compromised within my papers. In fact, the general
formulation of the theory I have used in \cite{Christian}
has been with us since the very dawn of the hidden variables program \cite{von}.
It is of course admirable to seek complete clarification of the fundamental issues raised within my papers, but it
is quite deplorable to do so by branding one's own misconceptions and prejudices as ``elementary errors'' within them.

\smallskip

{\underbar{\bf Note added in response to version 6}:} In this version of his preprint the shifting focus of Tung Ten Yong's
critique has shifted yet again. Apparently, I have once again misunderstood what his main objection is. Apparently, no
sooner have I made the mathematical consistency of my model explicit for his benefit, it is no longer the focus of his
objection. Apparently, the focus of his objection has all along been the operational adequacy of my model. But if so,
then I would simply say this: go back and read my papers---especially Sec. II of Ref.\cite{further}---before criticizing
them on the basis of misconceived ideas. For, not only has he failed to comprehend the operational details of my model,
but has also failed to show the basic understanding of what a proposed hidden variable model is required to do.
Since the operational requirements of such a model are quite well known and have been carefully analyzed by von
Neumann and others, and since I have shown elsewhere how exhaustively my model satisfies these requirements \cite{further},
I will not repeat them here. Suffice it to say that what is presumed by Tung Ten Yong as required of such models is at
best a gross exaggeration of what is {\it actually} required \cite{further}. One is of course quite free to differ in
opinion about such matters, but it is preposterous to herald such a difference of opinion as a ``key mathematical error'' 
in my paper.

\smallskip
 
Even more objectionably, in the last two paragraphs of this version Tung Ten Yong continues to make fallacious assertions
about the details of my model reflecting on his own misconceptions. Instead of patiently correcting his errors one after
another as I have been doing, I am tempted to leave these latest ones to the judgement of the readers. However, since I
am swimming against the paradigm, it is best that I continue correcting them, regardless of how tediously elementary they
are. The first one of his errors is in the statement: ``But from his equations it is obvious that the functionals,
curiously, can only obtain the zero multivector, and cannot obtain other scalar values within the domain.'' This
statement is simply false, as can be readily seen by comparing the Eqs.${\,}$(18) and (19) of my preprint \cite{Christian}
with the Eqs.${\,}$(\ref{codomaincliff}) and (\ref{codsuch}) above. Such carelessness in observation has been quite
emblematic of Tung Ten Yong's critique, and serves only to invalidate his arguments. What is more, he yet again
resorts in this paragraph to arguing in terms of changing the distribution ${{\boldsymbol\rho}({\boldsymbol\mu})}$
by itself, without any concurrent considerations of the local beables. The meaninglessness of his arguments based on such
a dubious form of reasoning has already been exposed above.

\smallskip

The next of Tung Ten Yong's errors is exhibited in his insistence that a ``directed measure is not the usual valid
probability
measure, since the former normalizes to a trivector (as in [my] equation (\ref{normscal})) but the latter normalizes to
a scalar one.'' This insistence is flabbergasting, especially after the pedagogical details I have provided above. Once
again this reflects his inability to comprehend a very simple manipulation within an elementary part of the geometric
algebra. In fact, the falsity of his insistence---that a ``summation of trivectors'' would yield a ``trivector'', and
not a ``scalar''---can be readily exposed by an elementary example taken from geometric algebra. Consider, for instance, 
a volume enclosed by a closed surface within the Euclidean space ${\mathbb{E}_3}$. It is a well known consequence of
the fundamental theorem of calculus within geometric algebra that the directed area of any closed surface vanishes:
\begin{equation}
\oint\;d^2{\bf x}\,=0,\label{Stokes'}
\end{equation}
where ${d^2{\bf x}}$ is the directed area element on the surface. This is of course nothing but an elementary consequence
of the familiar divergence theorem, but what it shows is that on any closed surface in ${\mathbb{E}_3}$ directed area
elements occur in pair with opposite orientations, which cancel when added, giving a scalar, namely zero (for a more
complete discussion see Ref.\cite{Hestenes}). Moreover, the above result easily generalizes to
arbitrary dimensions for the same reasons, giving
\begin{equation}
\oint\;d^m{\bf x}\,=0,\label{n-Stokes'}
\end{equation}
which exhibits how volume elements (or directed measures) in arbitrary dimensions can also add up to a scalar. Another
way in which operations between multivectors can give rise to scalars is through the geometric product, since this
product is invertible (unlike the products of vector algebra). My Eq.${\,}$(\ref{norm}) above is simply a particular
instance of these general facts. It shows how the product of two directed volumes with opposite orientations can end up as
a scalar, namely one: ${(+I)(-I)=1}$. The triviality of this elementary fact just about sums up Tung Ten Yong's critique.

\end{document}